\documentclass[aps,twocolumn,showpacs,preprintnumbers,prb,amsmath,
amssymb,amsfonts,superscriptaddress,floatfix]{revtex4}

\usepackage{amsfonts}
\usepackage{amssymb}
\usepackage{amsmath}
\usepackage{ulem}
\usepackage{color}
\usepackage{latexsym}
\usepackage{graphicx}
\usepackage{subfigure}
\usepackage{graphics}
\usepackage{floatflt}
\usepackage{epsfig}
\usepackage{overpic}
\usepackage{soul} 
\usepackage{varioref,xr-hyper} 
\usepackage[latin1]{inputenc}
\usepackage{mathbbol}
\usepackage{bbm}

\newcommand{\be}{\begin{equation}}
\newcommand{\ee}{\end{equation}}
\newcommand{\bea}{\begin{eqnarray}}
\newcommand{\eea}{\end{eqnarray}}

\def\lb{\label}

\newcount\bozza \bozza=0
\ifnum\bozza=1
\newdimen\shift \shift=-1.0truecm
\def\lb#1{%
{\label{#1}\rlap{\kern\shift{$\scriptstyle#1$}}}}
\else\def\lb#1{\label{#1}} \fi

\begin{document}
\title{Landau model to illustrate the process of learning and unlearning of nociplastic pain}

\author{Bel\'en Valenzuela}
\affiliation{Instituto de Ciencia de Materiales de Madrid, 
ICMM-CSIC, Cantoblanco, E-28049 Madrid, Spain}

\date{\today}

\begin{abstract} {
Recent advances in the comprehension of the consolidation of nociplastic pain points to a complex nonconscious learnt process of threat perception. Neurobiological education is emerging as a promising approach to unlearn nociplastic pain supported by biopsychosocial tools (exposition to movement, mindfulness, sharing group format...). However this approach is still poorly known by clinisians and society in general forming a communication problem that, unfortunately, perpetuate the suffering of the patients. We propose a Landau model to describe the process of learning and unlearning nociplastic pain to help to clarify this complex situation and facilitate communication between different sectors of society. Nociplastic pain corresponds to a first order transition with attention more likely in the alert-protection state than in the trust-explore state. Two appealing results of the model are that the perception of the critical context depends on the personal history about the symptom and that biopsychosocial loops are formed when there are alarming learnt historic information about the symptom together with confused and contradictory expert information as in nocebo messages. Learning and unlearning in the model correspond to a change in control parameters able to weight more alert-protected state, trust-explore state or neutral state. This description makes clear why neurobiological education is the ground therapy from which others must be built to embody the pertinent, clear and trustful information. 
}\end{abstract}

{FaceTime\bf \maketitle } 
 
Chronic pain is increasing at an alarming rate in recent years as exemplified by low back pain\cite{Maher:2017aa,Hartvigsen:2018aa,Rheumatology:2023aa}. Musculoskeletal chronic pain has been identified as one of the leading cause of disability worldwide\cite{Diseases:2020aa}. In addition, musculoskeletal conditions may increase the risk of other chronic disease such as cardiovascular disease, cancer and diabetes.\cite{Williams:2018aa} This disturbing situation has increased the research interest for a better understanding of chronic pain. Recently, nociplastic pain has been defined as a large component of chronic pain not associated to tisular damage\cite{Fitzcharles:2021aa}. This term was necessary because most chronic pain is non specific and it does not correspond to an underlying pathology.\cite{Hartvigsen:2018aa,Maher:2017aa}  From advances in cognitive and phenomenological science, there is strong evidence that the consolidation of nociplastic pain is a complex nonconscious learnt process of threat perception which can be formed by expectatives and/or learnt habits, alarming information from clinicians or experts and interpretation of the context giving rise to maladaptive loops\cite{Barrett:2015aa,Wiech16,Blasini:2017aa,Ongaro_2018}. This insight opens the possibility to overcome or attenuate nociplastic pain if it were possible to unlearn these beliefs and habits via the plasticity of the nervous system reducing the perception of threat. Biopsychosocial models\cite{Engel:1977aa,Loeser:2006aa,Gifford:1998aa,Vlaeyen:2012aa,Sullivan:2008aa,Gatchelbiopsycosoc2007} rooted in neurobiological education of pain (NBE) have emerged as the ground approach to help to this problem\cite{Nijs:2017aa,Explainpain,Lumley:2019aa,Arturosapiens}. 

Unlearning nociplastic pain via learning neurobiological education is not an intellectual learning but an embodied learning, meaning that it is a process where, in a safe and caring environment, the information is getting interiorized from the conscious patient to the nonconscious organism until it becomes the automatic perception. Thus, the patient makes sense of his/her own experience and develops an internal compass to discern what is threat, what is not and what is uncertain. When reducing the perception of threat the intensity of symptoms decreases, the person becomes more funcional and symptoms eventually might disappear. This is not an easy task since the patient with nociplastic pain presents a nonconscious learnt suffering pattern with intricate cognitive, emotional, attentional, motivational, motor, behavioral and social loops\cite{Bushnell:2013aa}. Physiologically, the entire nervous system including the brain, the endocrine system, the immune system and even the microbiota are taking part in the perception of threat\cite{Colombetti:2019aa,Rabey:2022aa} and both innate and adaptive immune responses modulate pain perception and behavior\cite{Murray:2021aa}. Therefore the process of interiorizing the information of neurobiological education might be different for each patient and might take different time. Since there is a threat perception of the organism this learning might require building a safe and caring social environment for the patient simultaneously with the active coping of the patient with their own recovery. That's why it is usually complemented with other techniques that adapt to needs and preferences of the patient to embody the information\cite{Rheumatology:2023aa} that we will call biopsychosocial tools. Examples of these tools are mindfulness\cite{Diez:2022aa}, exposition to movement, sharing group format, playing, imaginative analgesia, psychological assistant etc. Very positive effects of this therapeutic approach have been reported in migraine,\cite{Aguirrezabal:2019aa,Migra19} in musculo-skeletal pain,\cite{Galan-Martin:2019aa} fibromialgia\cite{Ittersum:2014aa,Barrenengoa-Cuadra:2021aa,Areso-Boveda:2022aa}. The advantages of this approach to pain are enormous since the patient is less exposed to secondary effects of pills. Most prescription pain killers cause significant side effects such as addiction and pharmacotherapy remains suboptimal,\cite{Deyo:2009aa} especially in the face of high placebo effects\cite{Ongaro_2018,Arandia:2021aa}. Finally, this embodied learning helps to understand that hypervigilance, anxiety, depression, anger, fear and catastrophising in the pain experience is part of the process and that other annoying symptoms the patient might suffer, such as insomnia, brain fog, ruminating thoughts, tense jaw, restless legs, tense muscles, digestive disorders etc also might belong to threat perception\cite{Gatchelbiopsycosoc2007}. 

Despite the advantages of the biopsychosocial framework this model has not yet permeated broadly neither to the health system nor to the society. The reason is complex and we will mention some aspects. First, pain is in the process of being understood with the definition of pain still changing.\cite{Cohen:2018aa,Apkarian:2019aa} Pain is also addressed at several levels from biochemistry to physiology, psychology, sociology and philosophy, each level with their own complexity and terminology. In addition, the enormous amount, complexity and generation of the research information about pain makes difficult that the essential information permeates to different sectors of the scientific community, to the clinicians and finally to the society. This creates a communication problem at different levels reflected in a lack of update of advances in the understanding of pain neither to graduate studies\cite{Watt-Watson:2009aa} nor to the expert community. At the end of the chain, this uncertainty is translated to the patient increasing his/her threat perception that fuels the pain. In addition, these patients are visiting several experts, clinicians or not, trying to understand her/his different symptoms increasing the already confused state. We will name all these experts and clinicians the expert culture, a name borrowed from Arturo Goicoechea\cite{Arturosapiens}. 

In this situation it is not surprising that the neurobiological education proposal is poorly known.  Unfortunately, this fact facilitates that patients absorbs erroneous beliefs many of them adopted from the expert community
\cite{Benedetti:2002ab,Blasini:2017aa,Setchell:2017aa}. Common misconceptions translated to the patients are "pain is related to tisular damage", or "the sensation of pain is proportional to tisular damage". Another issue are fragility messages such as "you have pain because your muscles are weak"\cite{Stilwell:2019ab}. These erroneous messages, so called nocebo effect, precipitate the consolidation of persistent pain\cite{Manai:2019aa}. This is even more important due to the bias of the mind to nocebo messages\cite{Baumeister:2001aa}. The uncertainty exposed above together with these misconceptions form a larger social loop where the patient is embedded. One could think that in these cases pain is not formed by maladaptive loops of the patient since the loops of the patient are adapted to his/her misinformed social milieu. This is in line to new definitions of health with emphasis in that the organism adapts to the interiorized biopsychosocial information\cite{Sterlingbook}. In this sense, these biopsychosocial loops are then adapted to society but maladaptive to life since there is a non necessary suffering pattern.

Implementing the biopsychosocial model is challenging.  It seems that a new curriculum in pain is not enough to prepare medical students but that it is essential both competence and compassion toward their patients.\cite{Murinson:2013aa} Since pain is related to a threat perception, conscious or not, to be in a trustful environment where the patient does not feel judged but listened, believed and understood is the starting point to initiate the embodied learning of NBE. The biopsychosocial model is also vaguely defined and there is the tendency to separate the patient into three domains (biological, psychological and social)  without taking into account the experience of the patient.\cite{Wideman:2019aa,Stilwell:2019ab}. As pointed out by Peter Stillwell and Katherine Harman\cite{Stilwell:2019ab}, to explain pain sometimes is used a reductionist approach where in the patient education the clinicians might use problematic pain explanations as "pain is in the brain" which is confusing to some patients who think they might have something wrong in their head or pain is not real but psychological\cite{Stilwell:2019ab}. In \cite{Stilwell:2019ab} it is proposed instead, understanding the subjective experience of the patient from the Enactive approach\cite{Varela:2016aa,Di-Paolo:2017aa}. 
The enactive perspective is a branch of embodied cognitive sciences based on dynamical systems, phenomenology and organizational approaches to biology. It aims to build a bridge between life and mind, investigating organisms embedded in their physical and social context. In this approach cognition is defined as 'sense-making', the capacity of an organism to evaluate different possible options and act in an adaptive manner to maintain and expand life.

 On the other hand, expert community and patients are skeptical about the proposal that pain can be learnt unconsciously and can be unlearnt learning about neurobiology of pain. In fact, it is indeed remarkable that embodied education in neurobiology can be of such enormous help for the well being of the person. For this, the consensus of the messages by the expert community is key to build trust.

In Ref.~\cite{Granan:2016aa} it was proposed that approaches from the adjacent field of Statistical Physics, that allows to model phase transitions, was the appropriate framework to understand chronification of pain and could be used as a communication tool. The idea put forward was to build an Ising model for positive and negative biopsychosocial factors relevant to pain although the model was not formally formulated. We also think that the analogy to phase transition is useful to illustrate the essential understanding of chronification of pain although instead of focusing on positive and negative biopsychosocial factors from an external perspective we propose to start from the subjective experience of the person. This will allow to also point out how is possible to unlearn the perception of threat in nociplastic pain. We prefer the phenomenological Landau approach\cite{landau2013statistical} to phase transitions to start with because it helps to discern the essential variables and parameters. It is also simpler, what makes it more attractive as a communication tool in diverse disciplines and to different sectors of society. Moreover, it is possible to connect Landau models with Ising models where the Ising models are the microscopic version of Landau models\cite{Altland:2010aa}.

In this article, we model  the automatic perception which can be either in an alert-protected state, a trust-explore state or a neutral state. This is determined by the following parameters of embodied information: information from senses about the context, the nonconscious or conscious historical experience related to the symptom and the information from expert community that polarize patient's opinion. The equivalent of the free energy\cite{landau2013statistical} is the patient sense-making.  This is a term borrowed from the Enactive approach\cite{Di-Paolo:2005aa}. Automatic attention is located in the most likely state in the sense-making landscape. Several sense-making landscapes corresponding to different subjective experience arise depending on the parameters: Zen, uncertain, hypervigilance, catastrophizing, curiosity, communicative... As a result it is seen that: 1. The critical context from where the alert-protected or trust-explore states arises depends on the personal history related to the symptom, this agrees with recent knowledge in neuroscience\cite{Lange:2018aa} and 2. A hysteresis loop is formed with the personal history and contradictory or misguided expert information. This hysteresis loop corresponds to the biopsychosocial loops found in patients\cite{Nijs:2017aa,Explainpain,Lumley:2019aa,Arturosapiens}.  The model is used to illustrate the nonconscious learning process of nociplastic pain with nocebo messages and the embodied learning of neurobiological education to dissolve the biopsychosocial loop. The model might help to communicate the synthesized information with a common thread and guide practitioners and health policies. It might also facilitate that the patient could make sense of his/her own experience. It can also be a tool to disseminate the benefits of a meaningful and updated biopsychosocial integrated framework to the society.  

In the following we present the derivation of the Landau model for the automatic perception. Next, we span the different sense-making landscapes. We also show the formation of hysteresis loops with expert information and historic information. We illustrate the process of learning/unlearning nociplastic pain using the model and we end up with a discussion and conclusions.

\section{Derivation of Landau model of the automatic perception}

Landau models\cite{landau2013statistical} were originally proposed to describe phenomenologically phase transitions common in nature where a control parameter varies: for example how iron is magnetized when lowering the temperature below a critical temperature or when increasing a magnetic field. Magnetization would be the order parameter which is zero above the transition temperature and different from zero below the transition. The temperature and the magnetic field are control parameters that when varied can make a transition from one state to the other state. The free energy is a functional of the control parameters and the order parameter whose minima determine the most stable states. The representation for a given set of parameters is a free-energy landscape with minimum points that correspond to the most likely states and will determine the state of the system. In the case of magnetization there would be three possible states: downwards magnetization, upwards magnetization and neutral. An influential and inspiring article by Phil Anderson\cite{Anderson:1972aa} in the context of condensed matter physics proposed that the concept of phase transition could help to understand emergent phenomena from interacting components at each hierarchical level of science, including life and mind. 

Most common phase transitions are of first or second order in the Landau classification \cite{Binder:1987aa}. In first order transitions there is a mixed state at the transition. For example, in the case of magnetization there would be a mix between upward and downwards magnetization. In second order phase transition there is however criticality at the transition. A very important concept which is related to large scale cooperative phenomena. In the case of magnetization at the critical transition all the magnetic moments cooperatively align in either upward magnetization or downwards magnetization and the magnetic susceptibility diverges. Extensions of the concept of criticality are widely used to describe life systems\cite{Bakcrit96,Langton:1990aa,Longo:2011aa} and neural activity\cite{Munoz:2018aa,OByrne:2022aa}. Psychodynamic processes have also a long tradition in dynamical systems\cite{Kelso:1995aa,Timmerman:2009aa,Barandiaran2013} and Landau theory is also used to model other subjective experiences\cite{Solnyshkov:2022aa}. 

Now we proceed to use the Landau framework to build up a model to delineate the process of learning/unlearning nociplastic pain. For that, we need to address the threat perception of the symptom. We will use inputs from phenomenologist and cognitive sciences about perception of pain or other symptoms related to an alert-protected state. It is not the scope of this work to achieve a comprehension of the complex process of perception of a sensation, we just borrow some intuitive concepts from the scientific literature to present the phenomenological model. 

Let's start by the sensation. We understand pain and symptoms as persistent sensations. Sensations are experienced throughout the day reporting about demands or needs from homeostasis and allostasis.\cite{Barrett:2015aa,Sterlingbook} Thus, they are a nonconscious evaluation of the needs of the organism. Physiologically, the information needed for the evaluation is circulating through the neuro-inmune-endocrine plus microbiota system and includes cognitive-emotional information from own history, context and culture. This forms a pattern of intricate rules aiming to survive and expand i.e. the process of homeostasis and allostasis. The sensation is expressed in our consciousness and urge us to interact with the external world to satisfy the need as an automatic response. For example, the hunger sensation urge us to look for food. We perceive the sensation, the evaluation that the sensation is hunger and the motivation to go for food. Consciously we can decide if we go for food or not. To describe a sensation are needed valence and arousal. Valence is related with how the organism validates the sensation if pleasant (positive) or unpleasant (negative). Arousal measures the intensity of the sensation, if modulation is low it is felt quite and, if it is large, it is felt agitated. Zero arousal corresponds to a neutral sensation.  

The automatic perception of the symptom, labeled by $\phi$ is the order parameter of the Landau model since from all the information available, it collects just the most relevant. We will understand this automatic perception as a semiconscious cognitive-emotional evaluation of the symptom where the historical, sensorial and expert information is integrated to discern the evaluation-motivational state: either alert-protection or trust explore\cite{Barrett:2015aa,Lange:2018aa,damasio2021feeling}. By semiconscious we mean that it is possible to become aware of this semiconscious evaluation by self-observation as in metacognition. The perception of the symptom  $\phi$ equals zero means the symptom is evaluated as neutral and there is no need. If $\phi \ne 0$ there is uncertainty either because there is a novelty or an inconsistency or a contradiction in the information perceived. Information reduces uncertainty so a cognitive-emotional causal query looking for sense arises in the default mode of the mind to remind intrinsic (from own history) or look for extrinsic information related to the symptom (from the context and social milieu). The evaluation can result in $\phi$ negative meaning the sensation is possibly dangerous for survival and protection is needed; or $\phi$ positive, corresponding to liveliness perception since there is trust that it is safe to explore. 

Naively one would think that when the valence of a sensation is negative (unpleasant sensation) then $\phi<0$ especially if arousal is large, but it is also possible that an unpleasant sensation will fade away to neutral state eventually as, for example, in the case a person has done exercise and feels stiff muscles. Internally is reminded previous experiences and other people's experiences with stiff muscles after sport. The automatic perception resolves that the sensation is known and will disappear, no alarm is sent to the individual, no much attention is given to the stiff muscles and at some point, $\phi$ might turn neutral and eventually arousal of the sensation turns to zero. 

Thus, there are several layers of evaluations. The nonconscious evaluation from the organism expressed in the consciousness by the sensation. The automatic perception of the sensation $\phi$, and the agent perception which, in principle, is a reevaluation of the perception and sensation in the present social and physical context to discern if following the automatism or not. These layers of evaluations might be wrong or contradictory. For example, in nociplastic pain, when the agent wishes to do his/her daily task, the organism evaluates pain and alert-protection perception and the agent cannot perform the task. That is, it is not possible for the intention's agent to become an action, agent and organism are not aligned. Alignment can come back consciously embodying NBE information. We will not model this feedback between the agent and the organism, just the automatic perception of the symptom from which the learning/unlearning process can be understood.

Having defined the 
automatic perception as the order parameter, $\phi$, we are ready to build up the Landau model.  In Statistical Physics, $F$ is the free energy and a potential minimum is the most likely state of the phase space of a system, i.e. states with lower potential corresponds to higher probability. Analogously, a hill denotes an unstable state. The potential landscape will change shape at the transition. In the present case the analogous to the energy is the sense-making $S$, a term borrowed from the enactive approach\cite{Varela:2016aa,Di-Paolo:2017aa}. As we have already mentioned, the perception of a sensation leads to a search for sense reminding intrinsic information (from own history) or extrinsic information (from physical or social context) related to the symptom. What makes more sense to survive or to expand is what determines the more likely state of the perception $\phi$ among the possibilities. The minus sign comes because the higher the sense-making, maxima in the landscape, the higher the probability.  To make analogy to Landau theory we prefer to add a minus sign in such a way that minima corresponds to likely states $-S=F$. Having this in mind we will call the different landscapes the sense-making landscapes. For that, we express $F$ expanded in powers of the order parameter $\phi$:

\begin{equation}
-S=F=-h_{ext}\phi+\frac{a}{2}\phi^2+\frac{h_{int}}{3}\phi^3+\frac{b}{4}\phi^4 \label{eq:model}
\end{equation}

In this expression all control parameters, $h_{ext}$, $h_{int}$, $a$ and $b$, are nonconscious embodied information of causal relations to infer perception of the symptom, i.e., meaningful information for survival or living concerning the symptom.  $h_{ext}$ denotes an external bias provided by the information from the expert culture about the symptom. To model the present situation of threat perception in nociplastic pain exposed in the introduction, $h_{ext}>0$ corresponds to precise and updated neurobiological information and $h_{ext}<0$ corresponds to confuse and nocebo messages when there is unjustified alarming information.  As we have already mentioned the expert advise has strong relevance since it is the one that can answer the uncertainty about patient's health. Next, $h_{int}$ is the symptom historical information enclosing previous learnt rules such as beliefs and expectatives from past experiences and learnt habits related to the particular symptom. $h_{int}$ is positive/negative corresponds to alarming/pleasant rules related to the symptom. Then, we define the parameter $a$ as it is common in Landau as $a=a_0 (T-T_0)$ where $T$ is the registered information by the exteroceptive and propioceptive senses  i.e. information about the actual context and the presence of the person in this context. $T_0$ is the critical value of the organism with innate stored rules about when the context becomes uncertain.
High $T$ means collecting abundant information from senses, low $T$ means collecting less information from senses and zero $T$ means no information from senses. $T$ is the only information not related to the symptom. Finally, $a_0$ and $b$ are innate positive parameters. By innate we mean the genetic tendencies of the person. We will comment on these two last parameters on the discussion section. %In Landau theory $a_0$ is the coefficient of the inverse of magnetic susceptibility and $b$ is the coefficient of the self-interaction between the magnetization. In the present context $a_0$ is the coefficient of the inverse of the sensitivity to expert culture and $b$ represents a measure of the capacity of metacognition i.e. the capacity of perceiving own automatic perception.  

\begin{figure}[thb] 
\includegraphics[width=0.9\linewidth,clip=true] {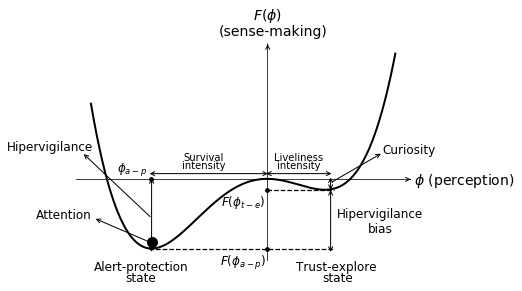}
\caption{ Typical sense-making landscape $F(\phi)$ versus the perception of the symptom $\phi$ showing deepest minima in the alert-protection state than in the trust-explore state. The black point is attention which is located in the deepest minima. Hypervigilance, curiosity, hypervigilance bias sense-making and pleasant/unpleasant intensity in perception are depicted. Negative perception defines survival and positive perception liveliness.}
\label{fig_Sensemaking} 
\end{figure}

In Fig.~\ref{fig_Sensemaking} we represent a typical sense-making landscape $F(\phi)$ with two minima that allow to define useful vocabulary to include all previous concepts.  Similarly the sensation has a valence and an arousal, the perception of the sensation, can also be positive, neutral or negative. Its intensity corresponds to the absolute value of the perception $|\phi|$. $\phi$ negative represents a survival perception and $\phi$ positive, a liveliness perception. The minima denote the most likely perception. The minimum in the survival perception is called the alert-protected state ($\phi_{a-p},F(\phi_{a-p}))$ and the minimum in the liveliness region, the trust-explore state ($\phi_{t-e},F(\phi_{t-e}))$. $\phi_{a-p}$ is the survival perception at the alert-protected state and $\phi_{t-e}$ is the liveliness perception at the trust-explore state. The sense-making value at the alert-protected state is called hypervigilance $F(\phi_{a-p})$. Since there is an alert-protected state, what makes sense is to look for information regarding the danger. On the other hand, the sense-making value at the trust-explore state is named curiosity $F(\phi_{t-e})$. When there is a trust, perception of the sensation in a safe environment, makes sense a natural curiosity to know about what is around.  We also define two bias, the perception bias:  $\Delta \phi=|\phi_{t-e}|-|\phi_{a-p}|$ as the difference between the intensity in the trust-explore state with respect to the alert-protection state and  the sense-making bias defined as the distance between the two minima, the trust-explore state respect to alert-protected state $\Delta F=|F(\phi_{t-e})|-|F(\phi_{a-p})|$. A positive bias in perception $\Delta\phi>0$ is optimistic and a negative bias pessimistic $\Delta\phi<0$. A positive bias in sense making $\Delta F>0$ represents curiosity bias and a negative bias in sense making $\Delta F<0$, hypervigilance bias. Attention is represented as a black point. If it is automatic it is more likely in the global minimum. Conscious attention can be in any extreme of the sense-making landscape depending on the person's will although might require more effort depending on the bias size. We finally define $\phi_{dm,a-p}(T=0)$ and $\phi_{dm,t-e}(T=0)$, not shown in the figure, that corresponds to a saturated perception where the mind is in complete default-mode in either the alert-protection state or the trust-explore state. The saturated perception appears when there is no information from senses $T=0$ or there is some $T\ne 0$ but there is enough $h_{ext}$ such that $\phi_{dm,t-e}(T=0,h_{ext}=0)=\phi(T,h_{ext})$. This saturation perception will appear in the hysteresis loops representing biopsychosocial loops.

\section{Sense-Making landscapes}

\begin{figure}[thb] 
\includegraphics[width=0.9\linewidth,clip=true] {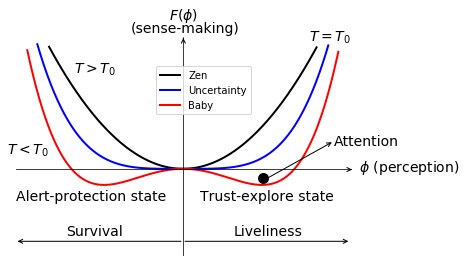} \caption{ Sense-making landscapes for $h_{int}=0$ and $h_{ext}=0$. $T_0$ is the critical context. For $T>T_0$ the most likely possibility is the minimum at the neutral state corresponding to the Zen landscape. At $T=T_0$ information from senses is equal to the critical context and corresponds to uncertainty landscape. When the information from senses is below the critical context, $T<T_0$, alert-protection and trust-explore states have equal sense-making value. This is the baby landscape. Attention is depicted as the black point in the trust-explore state. } \label{fig_Baby} 
\end{figure}

\begin{figure}[thb] 
\includegraphics[width=0.9\linewidth,clip=true] {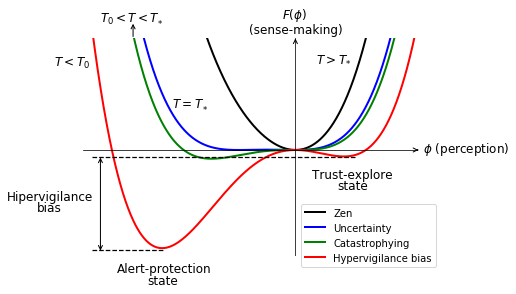} \caption{ Sense making landscapes for $h_{int}> 0$. $T_*=T_{0}+\frac{2h_{int}^2}{9a_0b}$ is the new critical context showing how the perception of the context depends on alarming previous rules. For $T>T_*$ there is the Zen landscape, $T=T_*$ corresponds to uncertainty with survival bias, $T_0<T<T_*$ to the catastrophising sense-making landscape and $T<T_0$ corresponds to hypervigilance bias. } \label{fig_Hypervigilance} 
\end{figure}

\begin{figure}[thb] 
\includegraphics[width=0.9\linewidth,clip=true] {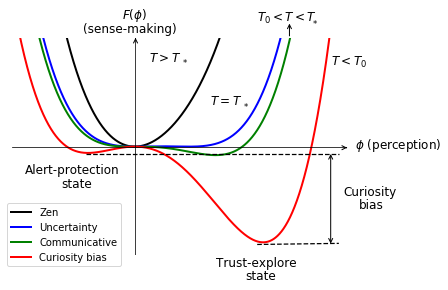} \caption{ Sense making landscapes for $h_{int}< 0$. The critical context $T^*$ depends now on comforting previous rules in $h_{int}$. Zen landscape for $T>T_*$, $T=T_*$ corresponds to the uncertainty with liveliness bias, $T_0<T<T_*$ to the communicative sense-making landscape and $T<T_0$ corresponds to curiosity bias. } \label{fig_Curiosity} 
\end{figure}

In the following we analyze different sense-making landscapes available in the model depending on different possible perceptions and sense-making values and bias. We identify the landscapes with mindsets in chronic pain such as hypervigilance and catastrophizing and with expansive states such as curiosity and communicative.  Notice that, different states also give rise to a particular social behavior that would be alert-protection-isolation and trust-explore-play.

Let's first analyze the simplest case with no expert information $h_{ext}=0$ and no historical information $h_{int}=0$. Fig.~\ref{fig_Baby} shows this escenario with three different landscapes. This is the typical free energy of a second order phase\cite{Binder:1987aa}. Since there is not previous information about the sensation what is perceived is what is felt and the sensation from the organism and the perception from the agent have the same valence and intensity. In this case $T_0$ is the critical context from a neutral state to an uncertain state. When the information from senses is bigger than the one focused on the critical context, $T>T_0$, there is a minimum at the neutral state $\phi=0$. The sensation is perceived as neutral. We will call this landscape the Zen landscape. Then, $T=T_0$ (blue line) is the critical value where uncertainty about the sensation sets in: the uncertainty landscape. At this value there are as many minima on the left than on the right interpreted as it is not known if the sensation is a threat or is safe. Then, below the transition $T<T_0$ in Fig.~\ref{fig_Sensemaking} (red line) there is less information from senses to focus perception on the sensation.  The landscape corresponds to a balance between alert-protected state (left-minimum) and trust-explore state (right-minimum). The state is balanced in the sense that the sense-making at the minima are equal $F(\phi_{a-p})=F(\phi_{t-e})$. There is not perception bias, $\Delta \phi=0$, and no sense-making bias, $\Delta F= 0$. We will called this landscape the baby landscape. Attention is represented as a black point. Thus, if the baby feels afraid the attention is on the left minimum $\phi_{a-p}$ and if the baby feels save and willing to explore, the attention goes to the right minimum $\phi_{t-e}$.  

Next, let's take $h_{int} \ne 0$ i.e there are previous learnt rules about the sensation. To illustrate nociplastic pain we set $h_{int}>0$. We remind positive $h_{int}$ comes from alarming rules related to the sensation by the organism. The possible sense-making landscapes is represented in Fig.~\ref{fig_Hypervigilance} and it is the typical free energy of a first order transition\cite{Binder:1987aa}.  In this case $T_{0}$ does not correspond to the critical information from senses representing uncertainty, but $T_{*}=T_{0}+\frac{2h_{int}^2}{9a_0b}$. From this expression it is seen that if there are many rules related to the symptom i.e. $h_{int}$ big, there are more contexts that are evaluated as possible threat, i.e. $T_{*}$ big. This result agrees with studies in cognitive sciences\cite{Lange:2018aa} where it is observed that alarming beliefs ($h_{int}>0$) distorts the perception of how danger is the context. We call this blue landscape in Fig.~\ref{fig_Hypervigilance} uncertainty pessimistic bias with more minima on the left than on the right meaning attention can wander between all these minima. At $T>T_{*}$ (black line) we just have a minimum and this state corresponds to the Zen landscape, as we have explained above. Attention can just be in the neutral state. For $T_0<T<T_*$ the organism is just in an alert-protected state that we have assigned it to the catastrophizing landscape. Attention is in the alert-protected state. Here there is pessimistic bias, $\Delta \phi <0 $, hypervigilance bias $\Delta F <0$ and no curiosity $F(\phi_{t-e})=0$. At $T<T_0$ (red line) there is a mixed state again with pessimistic bias $\Delta \phi <0 $ and an hypervigilance bias $\Delta F <0$ but with some curiosity in such a way that attention is more likely to be in the alert-protection state than in the trust-explore state. In this example, hypervigilance bias means that there is a tendency to absorb alarmed messages about the symptom. Notice that to focus on just information related to the symptom means lower information from senses ($T$ lower). Therefore, this mixed state is called the hypervigilance bias landscape. 

Let's consider now the case with $h_{ext}\ne 0$. If $h_{int}=0$, the figure represented in Fig.~\ref{fig_Baby} (red line) will have lower minima in alert-protection or trust-explore depending on the sign of $h_{ext}$. If this case represents a baby, $h_{ext}$ would typically represent the parents that polarize the baby uncertainty. If $h_{int} \ne 0$ and focusing in illustrating the case of nociplastic pain, $h_{ext}$ is the information from expert culture with strong impact in reducing uncertainty. In this case $h_{ext}$ can polarize the perception of the patient. We remind that $h_{ext}<0$ denotes misinformed information by the expert culture and  $h_{ext}>0$ corresponds to updated expert information in relation to the knowledge of pain. Of course, in a general case $h_{ext}>0$ might be also misinformed information representing placebo effect but we stick to the first situation to describe the nocebo problem in nociplastic pain. A negative value of $h_{ext}$ favors the alert-protected state as in the case of $h_{int}>0$ shown in Fig.~\ref{fig_Hypervigilance}. The explanation of different states would be similar where in addition to historical alarming beliefs and maladaptive habits there is misinformed messages from expert culture and proposition of rigid habits. 

In Fig.~\ref{fig_Curiosity} we represent the case when $h_{ext}>0$ corresponding to an updated expert information and/or $h_{int}<0$ corresponding to safe and comforting learnt rules about trust in the organism. Again $T_0$ becomes $T_{*}=T_{0}+\frac{2h_{int}^2}{9a_0b}$, meaning that there is optimistic bias to perceive the surround at the critical context. In this case the landscape at  $T_0<T<T_*$ represents the communicative sense-making landscape where the person is willing to share her/his discoveries about how to recover from the symptoms. There is then optimistic bias, $\Delta \phi >0 $, curiosity bias, $\Delta F >0$, and no probability for threat $F(\phi_{a-p})=0$. $T<T_0$ (red line) corresponds to a mixed state but now the global minimum is in the trust-explore state and there are again both, optimistic bias $\Delta \phi >0 $ and curiosity bias $\Delta F >0$, with attention more likely in the trust-explore state than in the alert-protection state. 

In summary, if $h_{int}=h_{ext}=0$ we have a second order phase transition with three landscapes forming when the information from senses is decreasing:  Zen, uncertainty and baby landscapes. In this case the sensations and the automatic perception of individual have same valence and intensity and there is no bias. When $h_{int}$ is different from zero there are first order transitions. The critical information from senses to arrive to an uncertain state is $T_{*}=T_{0}+\frac{2h_{int}^2}{9a_0b}$, meaning the critical context depends on the historical rules respect to the symptom $h_{int}$.  This case presents mixed states with bias in the perception and in what it makes sense in that situation. If the bias in perception is pessimistic the landscapes found when decreasing information from senses $T$ are first a catastrophizing landscape with zero probability for the trust-explore state and then a hypervigilance bias to focus in information to protect oneself. On the other hand, if the bias in the perception is optimistic, when decreasing information from senses, there are the communicative landscape with zero probability for the alert-protected state and curiosity bias landscape. 

Which landscape is the most appropriated? The organism evaluates with the information that it contains\cite{Sterlingbook}. If the information is wrong there might be an error in the evaluation. Opportunities of potential well-being thus need to be investigated. In the following we will deep in an error in evaluation due to confused or erroneous messages from expert culture.

\section{Hysteresis loop from expert information as a biopsychosocial loop}
%\begin{figure}[thb] 
%\includegraphics[width=0.9\linewidth,clip=true] {Fig_loop_T.png} \caption{ $h_{int.}$ }%\label{fig_loop_T} 
%\end{figure}
\begin{figure}[thb] 
\includegraphics[width=0.9\linewidth,clip=true] {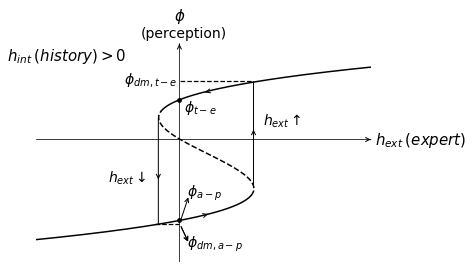} \caption{ Hysteresis loop between the perception of the symptom and the expert embodied information $h_{ext}$ related to the symptom. Dashed line do not belong to the loop because corresponds to the case where the polarization of the perception is opposite to the expert information. In the figure is observed that the loop is mostly in alert-protection since $h_{int}>0$ and the embodied neurobiological education $h_{ext}>0$ must have a big value to counteract the previous bias $h_{int}$. The default mode in the alert-protection state $\phi_{dm,a-p}$ and in trust-explore state $\phi_{dm,t-e}$ are depicted as well as the most likely perceptions $\phi_{dm,a-p}$, $\phi_{dm,t-e}$ at $h_{ext}=0$. $h_{ext\uparrow}$/$h_{ext\downarrow}$ represents the value of the embodied expert information when there is a transition from survival->liveliness/liveliness-survival perception.} \label{fig_loop_hext} 
\end{figure}
Imagine there are different criteria from the expert culture i.e. $h_{ext}$ varies for a given history of the person $h_{int}$ related to the symptom and context $T$. In the case of first order transitions with mixed states corresponding to the hypervigilance bias and to the curiosity bias sense-making landscapes, hysteresis loops are formed of perception $\phi$ respect to information absorbed from expert culture $h_{ext}$. In the hypervigilance loop there is a bias for pessimistic information and in the curiosity loop there is a bias for optimistic information. In the present case for nociplastic pain we will relate negative $h_{ext}$ with nocebo effect and positive $h_{ext}$ with neurobiological education. 

Mathematically the hysteresis loops are obtained 
 from the minimum of $F$ with respect to perception to find the most likely states:
\begin{equation}
\frac{\partial F}{\partial \phi}=0 \rightarrow h_{ext}=a \phi+h_{int}\phi^2 + b \phi^3
\end{equation}
%If $h_{ext}=0$ and disregarding the unstable solution $\phi=0$ for now we get $a+h_{int}%\phi+b\phi^2=0\rightarrow \phi=\frac{1}{2b}\{ -h_{int}\pm\sqrt{h_{int}^2-4ba_0(T-T_0)}\}$. %In Fig. is displayed the dependence of $\phi$ on $T$ where it is visualized the loop in T.

The hysteresis loop $\phi$ versus $h_{ext}$ in hypervigilance bias is represented in Fig.~\ref{fig_loop_hext}. 
In the figure is displayed the survival perception at the alert-protected state $\phi_{a-p}(h_{ext}=0)$, the liveliness perception at the trust-explore state $\phi_{t-e}(h=0)$ and the saturated perception corresponding to the default modes at the alert-protected state $\phi_{dm,a-p}$ and trust-explore state $\phi_{dm,t-e}$. $h_{ext \uparrow}$ is the absorbed expert information needed to go from alert-protection state to trust-explore state and $h_{ext}\downarrow$ is the absorbed expert information to change from the trust-explore state to the alert-protection state. In the case of nociplastic pain $h_{ext}\downarrow$ will correspond to the absorbed nocebo messages needed to change states from trust-explore to alert-protection and $h_{ext \uparrow}$ to the absorbed neurobiological education needed to change from alert-protection to trust-explore. 

The loop starts at $\phi_{t-e}(h=0)$. For negative $h_{ext}$ there are alarming messages and $\phi$ is decreasing towards $h_{ext}\downarrow$. The dashed line has a negative slope meaning that the trust/threat perception is opposed to information given by the expert culture $h_{ext}$ and then do not polarize the opinion of the patient: on the contrary, patient's opinion is opposite to expert information. Since we have started with the assumption that the expert information polarizes the patient's opinion we do not consider this case. Then at $h_{ext}\downarrow$ the perception turns to the default-mode of alert-protection $\phi_{dm,a-p}$. If now neurobiological education is it absorbed, the threat perception is decreased until finally reaches $h_{ext \uparrow}$ where there is a change in the state to $\phi_{dm,t-e}$. Then, there are nocebo messages and the loop starts again.

The meaning of this hysteresis loop is that the patient is confused by the contradictory information between nocebo and neurobiological education. We associated with biopsychosocial loop meaning bio in the symptom, psycho from perception and social from information from expert culture $h_{ext}$. This loop give rise to the cognitive, emotional, attentional, motor, motivational, conductual loops that help to consolidate the persistent symptom. This fact illustrates the necessity for the coherence in the information in the expert community, in media, at university and at school.

\section{Learning and unlearning persistent pain illustrated by the model}

 In the model unlearning nociplastic pain is just changing parameters: learnt nocebo messages $h_{ext}$ to trustful and updated neurobiological information $h_{ext}'$, alarming past learnt rules $h_{int}$ to new rules from revising meaning of previous ones, $h_{int}'$ and even training senses to collect more sensorial information $T'$ or training conscious attention to realize the no permanent character of the perception states. Clearly the patient is the main character taking an active coping in all the process to embody the new information. Let's illustrate this process of learning/unlearning chronic pain with an hypothetical example. 
 
Let's consider there is a perception $\phi$ of a sensation that it is an ache in the neck. If the neck-ache remains and there is uncertainty because the pain is new or disturbing,  the organism goes from a neutral state to an uncertainty landscape $T=T^*$, blue line depicted in Fig.~\ref{fig_Hypervigilance}. The default mode of the mind will be wandering with ruminating thoughts correlation-causal possibilities about the symptom. For example, do I have to worry about the neck-ache? I have been told that screens force a bad neck posture. Should I go to the doctor? Should I buy another screen/mouse?(attention is in the alert-protection state). Let's move a little bit or let's go for a walk (attention is in the trust-explore state). I think it is nothing to be worry about (attention is in the neutral state) etc. There might be the possibility that there is a bad memory about neck ache because there was some accident some years ago. In this case $h_{int}>0$ and $T^*=T_0+(2h_{int})^2/(9a_0b)$. Thus, the critical context is uncertainty with pessimistic bias in perception. The interest about the neck-ache increases and the patient, conscious or not, focuses on information about it, paying less attention to senses information, i.e. $T$ decreases. In principle, health experts have privilege information about pain and when the patient goes to the health expert expects to make sense of its pain and specially to get relief of its symptom.  Imaging experts infer from X-ray a cervical deviation $h_{ext}<0$ and tell the patient that pain arises because she/he acquire a bad posture when she/he works with computers. Nowadays it is well known that this information and recommendation can even worse the pain since a new fear about no correct position is leaking in the organism which feels the threat.\cite{Lederman:2011aa} The survival organism is in the hypervigilance landscape shown in Fig.~\ref{fig_Hypervigilance} (red line) with already alarming information about the neck and nocebo messages. Pain might consolidate, becoming persistent and sensitive to sitting on a chair i.e. context information included in $T^*$.  This experience disrupts the person's life since the organism finds danger at his/her workplace and other symptoms as brain fog and intrusive thoughts might appear when trying to concentrate at work giving rise to frustration. This will fuel the evaluation of threat, other symptoms corresponding to the alert-protection state might arise such as tense jaw, insomnia, digestive disorders... Each symptom will have their own sense-making landscape. The patient goes from one expert to other but no tisular damage is found. At this point the person is suffering and might distrust the expert community and distrust his/her organism.\cite{Kusch:2018aa}

Imagine now the person decides to visit a neurobiological education clinic. The information provided by the clinicians is different $h'_{ext}>0$. At the beginning there will be the biopsychosocial loop due to the contradictory information shown in Fig.~\ref{fig_loop_hext}. How can the person trust NBE if the most likely state is alert-protected state without trust in neither the expert community nor in his/her organism? Therefore, the first challenge is to build trust and maintain trust. The time needed to build trust will depend on the information embodied from own history $h_{int}$, from the experts $h_{ext}$ and from the context by senses $T$. That's why to build a safe and caring environment and a clear, accesible and honest information about pain is so relevant. Then, when trust is built between patient and clinician it becomes possible an active coping of the patient and a compromise to go through the practice. Certainly this trust building
 is necessary to start with but also during all the process of unlearning the threat perception when embodying learning NBE. 

Concurrently, the patient helped by the clinicians explores threat perception is not permanent playing with conscious attention, his/her sense-making landscape and how to infer nonconscious rules in $h_{int}$. How nonconscious rules can be identified if precisely there are not conscious? When learning about NBE there might be a contradiction between the new information and the person's misbeliefs. The contradiction might be disturbing and leads to the uncertainty state and the biopsychosocial loop. These contradictions can be also identified when listening the narrative of the patient, observing body language and behavior. It is necessary to approach to these contradictions gently and with empathy since if not alert-protection will emerge. Empathy might arise when the expert community realizes their own personal biopsychosocial loops maladaptive to life and understand how difficult it is to dissolve them. The patient can also become aware, via observing and exploring with curiosity instead of hypervigilance, the default mode of the mind and the own maladaptive cognitive, emotional, attentional, motivational, motor and conductual loops. From this exploration might be possible to infer misbelieves and maladaptive habits in $h_{int}$. In the present example, the patient will learn in NBE that many people with strong cervical deviation do not have any pain (correlation not equal to causality), will learn that there is no correct position but a position for each occasion, will also learn that all the symptoms arise from the threat perception which points to the root of the problem etc\cite{Nijs:2017aa,Explainpain,Lumley:2019aa,Arturosapiens}. All this new learning contrasts with previous expert information.  The updated information needs to be embodied exploring with curiosity for example by playing when the patient feels safe with any possible biopsychosocial tool available. Playing safely will also change how much information is extracted from the context, $T'>T$. Notice that playing might find some resistance because what the person wants it is to get rid of the pain. It is needed to remind that to embody the new  information it is necessary to explore without any objective, like a baby.  A challenging issue is that in the process the person might arrive to the catastrophizing state where the patient feels that there is no hope. However, notice that catastrophizing is closer to the neutral state than the hypervigilance state. We interpret this as the Fenix effect, where from the total suffering emerges a new perception when information from senses is allowed. Becoming aware of this state might be part of the process. In addition, resistance, pain and symptoms will reappear from time to time and all the unlearning process starts again but with a learnt base. Patience is necessary with trust in own organism. If finally the belief is dissolved, $h_{int}\to h_{int}'$, the sense-making landscape will change accordingly with more probability in the trust-explore state than before and less probability in the alert-protection state. The sense-making landscape can be also communicative where the patient is willing to tell his/her recovering experience. The clinician might become aware by a different narrative and a different body language. 

It is clear then, that the patient becomes the main character in his/her own recovery guided by experts in NBE and complementing with biopsychosocial tools adapted to the patient. The recovery time will be particular of each person. It might also happens that there is a remanent pain and relapses but the patient increases her/his functionality and thus her/his quality of life.

\section{Discussion and conclusions}

The present Landau model describes phenomenologically and qualitatively key aspects in the perception of a symptom $\phi$: 1. The contribution of personal history, physical context and expert culture to build the perception. 2. Optimization of the sense-making to discern if perception should be in an alert-protected state, in a trust-explore state or in a neutral state; 3. The automatic attention located in the deepest minima of the sense-making landscape and the conscious attention that could be located in any extrema of the sense-making landscape. 4. Second order transitions are derived if there are not past learnt rules, $h_{int}=0$, and first order transitions if there are past learnt rules, $h_{int}\ne 0$. 5. There are possible sense-making landscapes where different stages of the subjective perspective can be identified. For second order transitions these sense-making landscapes are Zen, uncertainty and baby and for first order transitions: uncertainty bias, hypervigilance bias, catastrophyzing, curiosity bias and communicative. 6. Unlearning corresponds to a change of the parameters from nocebo messages to NBE education $h_{ext} \to h_{ext}'$, changing meaning of learnt rules $h_{int}\to h_{int}'$ and training senses $T \to T'$. As a result, in first order transitions the critical context depends on the personal history of the person $h_{int}$. This agrees with neurosciences studies where it is found that the personal context is inferred by beliefs of the person\cite{Lange:2018aa}. Interestingly, from a different perspective, there have been proposals using neural networks to explain some mental illness as a disruptions of criticality\cite{Munoz:2018aa,OByrne:2022aa} which agrees with the view of pathology as a first order transition in this simplified model. In first order transitions we also find the formation of hysteresis loops interpreted in this work as a biopsychosocial loop. Hysteresis loops have also been proposed to explain perception in the context of neural representations\cite{Trapp:2021aa}. This model is applied to address the learning/unlearning process of the threat perception given in nociplastic pain and to highlight the nocebo effect in persistent pain as a biopsychosocial loop maladaptive to life. 

The result $T_*=T_0+(2h_{int})^2/(9a_0b)$ also points out that the extra term, $(2h_{int})^2/(9a_0b)$, besides the historic information of learnt rules, $h_{int}$, depends on the innate parameters $a_0$ and $b$. In statistical physics $a_0$ is related to the susceptibility to the magnetic field $\chi \sim 1/a_0$ and in the present model would be the perception susceptibility to expert information. Thus, for bigger $a_0$, there will be lower susceptibility to expert information and the extra term in the critical context $T^*$ will decrease, what makes sense. There might be people more sensitive to expert information ($a_0$ small) than others (big $a_0$). On the other hand, $b$ is related with self-interaction\cite{Altland:2010aa}, self-perception in the present case. If self-perception is interpreted as the perception of the perception it also makes sense that bigger the self-perception smaller $T^*$. Thus, for a given $h_{int}$, if there is not much sensitivity to expert information and there is a strong capacity of self-perception, the critical context is closer to the one in a second order transition $T \to T_0$. In the model $b$ cannot be very large because this would mean that other powers in perceptions would be necessary such as $\phi^6$. For a deep understanding of the consequences of these parameters and precise cognitive definitions a thorough study connecting the Landau model with Statistical Physics is necessary\cite{Altland:2010aa}. This will be left for future studies.

The model is simple and intuitive and clarifies why embodied neurobiological education goes to the core of the problem instead of just improving symptoms. It also points out the way to recover with an active coping of the patient. Neurobiological education helps to point out nocebo messages and other misconceptions and make sense of patient's experience. There are other approaches that aim to get rid of the symptoms but pain comes again since misconceptions remain and then the hypervigilance evaluation. That is, improving symptoms relieve the patient but do not change the landscape, just attention wanders from the alert-protected state to the metastable trust-explore state or to the neutral state while becoming aware of misconceptions and embodying the information with appropriated biopsychosocial tools do change the landscape.  The model also makes clear why is important that all clinicians share the same knowledge about pain. Finally, the model illustrates how NBE is extremely useful to prevent persistent pain and other symptoms.  

A clear limitation of the model is that the present Landau formalism is static and we introduce an effective dynamic by changing the control parameters reporting information from context, patient history and expert culture concerning the symptom. This dynamic does not correspond to time since each time there are different sensations and a persistent sensation is just more likely in time. The dynamic corresponds to a variation in the embodied information $\delta h_{int}$, $\delta h_{ext}$, $\delta T$ what will be reflected in a variation of the perception. Another limitation is that the model does not include the negative/positive feedback loop that will arise in the hypervigilance/curiosity bias landscape. A non-equilibrium model will be necessary to address this effect. This study will also properly account for the probability of the trust-explore or alert protection metastable states in the mixed state\cite{Binder:1987aa} what again requires the development of the model from first principles in Statistical Physics\cite{Altland:2010aa}. However, for the purpose of this work, that is, proposing a minimal model to facilitate communication, the model is enough to illustrate the problematic of learning nociplastic pain and how to unlearn it.

The model can be used to address other mental syndromes such as anxiety, depression, mioclonus symptoms, addictions etc which seems to have a common underlying mechanism.\cite{Barrett:2015aa,Arandia22} It is interesting to notice  that the biopsychosocial loops are in both, hypervigilance bias and  curiosity bias landscapes, as might happens with screen addictions where curiosity bias is looking for the sensation of surprise. Here the patient instead of avoiding the sensation as in pain, is looking for the sensation. The model could be also adapted to pathologies where the rules learnt by the organism opposes to the expert rules as for example in a maniac state. Anosognosia is common in mental syndromes and it is not that surprising that if it would be possible to become aware of misbeliefs and mishabits this might be of extreme relevance for the recovering of the patient.

Remind the patient's biopsychosocial loop is adapted to his/her embodied learnt rules. This is a different perspective of seen mental pathologies, including nociplastic pain, as dysregulated processes. This perspective motivated the definition of allostasis as stability through change to adapt to different needs of the organism.\cite{Sterlingbook} This requires prediction of the needs to satisfy them before they arise. Health is then define as the capacity for adaptive variation and disease as a compression of this capacity in contrast to the traditional definition of health as a list of "appropriate" lab values and disease as "inappropriate" values based in the control of homeostasis. The term allostatic load is used to refer to disease as a maladaptive loop behavior by the organism which is not dysregulated but coherent with their own innate and learnt rules. Allostasis thus enlarge the scope of health allowing to deal with cognitive and emotional symptoms. In this context, chronic pain has been described in terms of allostatic load.\cite{Borsook:2012aa,Rabey:2022aa}

In a long term processes however, the allostasis perspective "stability through change" might not be enough since in the historical process of life there is not stability but a continuous transformation where a process of individuation might emerge. This is in line to the proposal of extending criticality and symmetry breaking where the living state of matter is interpreted as an ongoing extended or critical transition always transient to a renewed organism.\cite{Longo:2011aa}  We conceive the learning process in the line to the proposal given in the Enactive plus Simondonian approach\cite{Arandia22} which emphasizes that "growth and transformation processes can be arguably be seen as fundamental for self-individuation for humans, not only subsistence". This devenir seems in line with the process of individuation proposed by Simondon as generation of metastable states by transforming  tensions to the environment or to the society\cite{Simondon:2020aa}.

\section{Conclusions}
We have built a Landau model to address the subjective perspective of a patient. The order parameter is the perception of a symptom and the control parameters are the context from senses, the embodied history and the embodied information from expert culture about the symptom. The model allows to show different perception scenarios corresponding to different sense making landscapes where automatic attention is placed in the most likely state. For second order transitions there are the Zen, uncertainty and baby landscapes. First order transitions present bias either for the alert-protected state or the trust-explore state giving rise to other possible landscapes: uncertainty bias, hypervigilance bias, catastrophyzing, curiosity bias and communicative. From the model is derived two interesting results well known in cognitive science : 1. the critical context where uncertainty appears depends on nonconscious misconceptions and mishabits about the symptom and 2. an hysteresis loop named, the biopsychosocial loop, arises in perception when there is confused expert information together with nonconscious alarming historical information. We apply this model to illustrate the threat perception given in nociplastic pain and the unlearning process via embodied neurobiological education. Learning and unlearning corresponds to changing control parameters, namely, a revision of  nonconscious misconceptions and mishabits, updated and trustful expert information and training senses and attention. 

From this model is clearly seen that the alarming increasing rate of chronic pain could be partly explained by nocebo and confused expert information that creates a threat perception in the patient and precipitate the organism into an alert-protection state. Within the embody learning of NBE the patient might identify these nocebo messages, investigate its own sense-making landscape and infer own alarming beliefs and mishabits.  Embodied learning of neurobiological education emerges as a valuable tool to reduce the perception of threat, prevent the chronic pain burden and antifragilize citizens who develops their own internal compass to be in the world. The strongest policy effort will be to promote this embodied neurobiological education besides clinicians, to the whole society from schools, to universities and media. This will avoid loops from nocebo effect, value the importance of the trust-explore state and of making sense of own experience.

\section*{ACKNOWLEDGMENTS}
 B.V. acknowledges deeply to Arturo Goicoechea and I\~{n}igo Arandia.

\bibliography{painwithoutdamage1}

\end{document}